\documentstyle[apb_cite,epsf,rotate]{mn}

\newcommand{\etal}{{\rm et al. }}

\newcommand{\Msun}{\mbox{M$_{\odot}$}}

\newcommand{\Mdot}{\mbox{M$_{\odot}yr^{-1}$}}

\title[The ``Outside--In'' Outburst of HT Cas.]
{The ``Outside--In'' Outburst of HT Cassiopeiae.}

\author[Ioannou et\,al.] { Zach Ioannou,$^{1}$\thanks{email:
zac@astro.keele.ac.uk} T. Naylor,$^{1}$ W.F. Welsh,$^{1,2}$ M.S. Catal\'{a}n,$^{1,3}$ 
W.J. Worraker,$^{4}$ \cr
N.D. James$^{5}$\\
$^{1}$Keele University, Department of Physics, Keele, Staffordshire,  ST5 5BG, UK\\
$^{2}$University of Texas, Department of Astronomy, Austin, Texas, 78712, USA\\
$^{3}$University of St Andrews, Department of Physics and Astronomy, North Haugh, St Andrews, Fife KY16 9SS\\
$^{4}$65, Wantage Road, Didcot, Oxfordshire, OX11 0AE, UK\\
$^{5}$11, Tavistock Road, Chelmsford, Essex, CM1 6JL, UK} 

\begin{document}

\maketitle

\begin{abstract} 
We present results from photometric observations of the dwarf nova 
system HT Cas during the eruption of November 1995. The data 
include the first two--colour observations of an eclipse on the rise 
to outburst. They show that during the rise to outburst the disc 
deviates significantly from steady state models, but the inclusion 
of an inner-disc truncation radius of about 4 $R_{wd}$ and a ``flared'' 
disc of semi-opening angle of $10^{\circ}$ produces acceptable 
fits. The disc is found to have expanded at the start of the outburst to
about $0.41R_{L1}$, as compared to quiescent measurements.  
The accretion disc then gradually decreases in radius reaching $<0.32R_{L1}$ 
during the last stages of the eruption. Quiescent eclipses were also observed 
prior to and after the eruption and a revised ephemeris is calculated.
\end{abstract}



\section{Introduction}

For many years there have been theoretical predictions that accretion discs can
undergo outbursts which start either close to their inner edge, or their outer
edge (Meyer \& Mayer--Hofmeister 1984; Mineshige \& Osaki 1985; Ludwig \& Meyer 1998).
In this and the following paper we use the eclipsing dwarf novae to resolve
spatially their outbursts, and prove that one outburst is ``outside--in'', the 
other ``inside--out'' \cite{Webb99}. This is the first time an ``inside--out'' outburst 
has been resolved, and should be viewed as a major vindication of disc instability theories.
In both cases we find that the discs have extensive vertical structure, which is shown
to be the cause of problems of interpretation in the only other observation of an
``outside--in'' outburst (Vogt 1983; Rutten \etal 1992).

The dwarf nova observed in this paper is the SU UMa star HT Cas, whose orbital period
is about 106 min. (Patterson 1981; Zhang, Robinson \& Nather 1986; Wood, Horne \& Vennes 1992)


\section{Observations \& Data Reduction}

\begin{table*}
\begin{minipage}{170mm}
\begin{center}

\caption{List of observations. Eclipse timings are in Barycentric Dynamical
Julian Date. The state of the HT Cas system is denoted with "Q" for Quiescence
and "O" for Outburst.}

\begin{center}
\begin{tabular}{ccccccccc}

Date        & Eclipse& BDJD     &   Observatory & Filter &State& Cycle time  & Depth & FWHM    \\ 
                                                                                              
            &        &2450000+  &               &        &     &   (s)       & (mag) & (phase) \\ 
                                                                                                         
            &        &          &               &        &     &             &       &         \\
                                                                                                       
1995 Oct 27 & 85415  &18.51448  &   Keele       &   V    & Q   &    20       &       &         \\ 
                                                                                                          
            & 85416  &18.58809  &   Keele       &   V    & Q   &    20       &       &         \\ 
                                                                                                        
1995 Nov 17 & 85700  &  ----    &   Keele       &   V    & O   &     8&-0.98 & 0.054${\pm}0.003$ \\ 
                                                                                                          
       &&39.50388  &   St.Andrews  &   R    & O   &    20       & -0.94 &0.061${\pm}0.003$\\ 
                                                                                                          
&85701  &39.57750  &   St.Andrews  &   R    & O   &    25       & -0.95 &0.054${\pm}0.003$\\ 
                                                                                                          
&85702  &39.65112  &   St.Andrews  &   R    & O   &    20       & -0.91 &0.059${\pm}0.003$\\ 
                                                                                                          
&85703  &39.72476  &   St.Andrews  &   R    & O   &    20       & -0.91 &0.058${\pm}0.003$\\ 

       &&  ----    &   Keele       &   V    & O   &     8       & -1.01 &0.056${\pm}0.003$\\ 

1995 Nov 18 & 85711  &40.38743  &   St.Andrews  &   R    & O   &    20&-1.21 &0.048${\pm}0.002$\\ 
                                                                                                          
       & 85712    &40.46079  &   St.Andrews  &   R    & O   &    20       &-1.33& 0.046${\pm}0.002$\\ 
                                                                                                          
       &85713  &40.53421  &   St.Andrews  &   R    & O   &    20       & -1.36&0.047${\pm}0.002$\\ 

       &85714  &40.60761  &   St.Andrews  &   V    & O   &    25       & -1.61&0.040${\pm}0.002$\\ 
                                                                                                          
1995 Nov 19 & 85724  &41.27144  &   Essex       &$Clear^{a}$& O   &  25& -2.23 &0.036${\pm}0.002$\\ 

       &85725  &41.34501  &   Essex       &$Clear^{a}$& O   &    25    &-2.42 &0.035${\pm}0.002$\\ 

       &85727  &41.49239  &   St.Andrews  &   R    & O   &    20       & -2.01&0.038${\pm}0.002$\\ 
                                                                                                          
       &85728  &41.56605  &   St.Andrews  &   R    & O   &    20       & -2.04&0.039${\pm}0.002$\\ 
                                                                                                           
       &85730  &41.71326  &   St.Andrews  &   V    & O   &    25       & -2.58&0.038${\pm}0.002$\\ 
                                                                                                           
1995 Nov 22 & 85769  &44.58551  &   Keele       &   R    & Q   &    25       &       &         \\ 
                                                                                                          
1995 Nov 29 & 85863  &51.50834  &   Keele       &   V    & Q   &    35       &       &         \\ 
                                                                                                         
1997 Jan 14 & 91454  &463.26995 &   Keele       &  Clear & Q   &    15       &       &         \\ 
                                                                                                     
       &91455  &463.34345 &   Keele       &  Clear & Q   &    15       &       &         \\ 

       &91456  &463.49075 &   Keele       &  Clear & Q   &    15       &       &         \\ 

1997 Mar 02 & 92093  &510.33020 &   Essex       &   V    & O   &    45       &-2.12 & 0.038${\pm}0.002$ \\ 

1997 Sep 05 & 94621  &696.51040 &   Keele       &  Clear & Q   &    25       &       &         \\ 

1997 Sep 10 & 94689  &701.51845 &   Keele       &  Clear & Q   &    25       &       &         \\ 

\end{tabular}
\end{center}
\protect\label{tab:observation_log}
\end{center}
\end{minipage}
$^{a}$\it{The unfiltered response for this CCD approximately resembles a broad V band}
\end{table*}

The observations of HT Cas, from the 1995 November eruption were carried 
out with the CCD system on the 0.95m James Gregory Telescope (JGT) at the University 
of St. Andrews (Bell, Hilditch \& Edwin \shortcite{Bell93}), the 0.6-m Thornton Reflector 
at Keele Observatory \cite{Somers96b} and with a 0.3m Newtonian reflector in Essex. 
Table~\ref{tab:observation_log} lists all the observations. Timings are in Barycentric 
Dynamical Julian Date (BDJD).

The raw images where processed in the standard way for the instruments in
question, as outlined in \scite{Somers96b} and Bell~{\etal}~\shortcite{Bell93}.
The data from Essex were dark subtracted and flat fielded in the standard way
for CCD photometry. The data were extracted using an optimal extraction technique by 
Naylor \shortcite{Naylor98}. The values for the visual magnitudes for the standard 
stars were adopted from \scite{Misselt96}. 

\begin{figure}
\vspace{7.5cm}          

\includegraphics{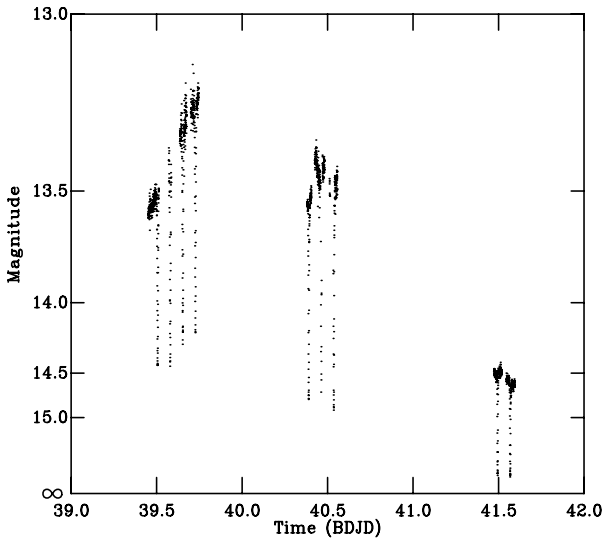}
\caption{Overall R band light curve of the November 1995 outburst. V-band
data have very similar flux levels. Position along the y-axis is proportional to 
flux.}
\label{fig:alllcvs}
\end{figure}

\begin{figure}
\vspace{7.5cm}          

\includegraphics{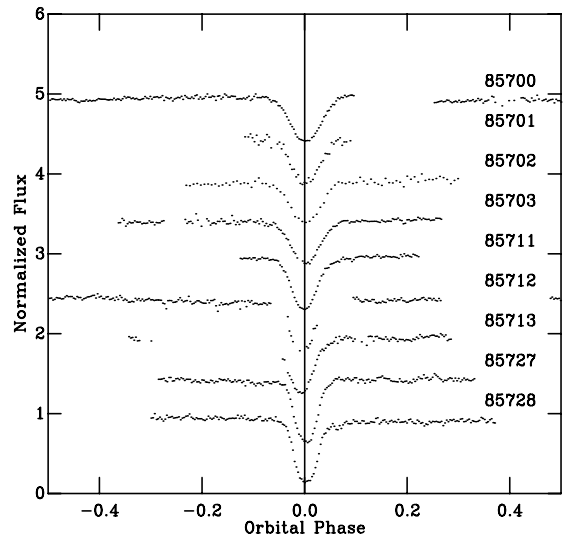}
\caption{The R-band outburst light curves from St. Andrews. Time resolution is 15 
seconds (eclipse 85701 is 20 seconds) with 5 seconds ``dead time''. Note the
drift from phase zero of eclipses 85712 and 85713.}
\label{fig:rband}
\end{figure}


\section{The Light Curves}

The overall outburst light curve is depicted on Fig.~\ref{fig:alllcvs}. The rise to 
outburst looks very steep with a linear decline. Our coverage of the outburst rise 
phase starts at a point where the system had reached a brightness level about 
3 magnitudes above its normal quiescent value. A steep rise is observed of less than 
30 hours between the last observed quiescent point (G. Poyner; priv. com.) and the peak 
of the outburst. The coverage of the decline phase was very good with the system being 
about one magnitude brighter than its quiescent value at the end of our coverage. 
The duration of this normal outburst is therefore estimated to be 3.5 to 4 days.

\begin{figure}
\vspace{7.5cm}          

\includegraphics{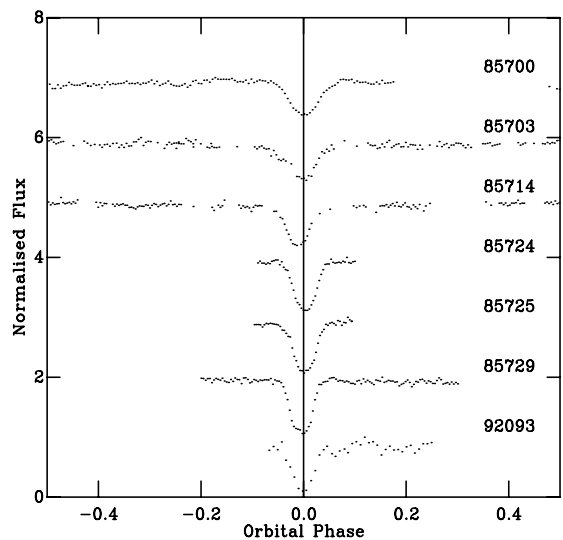}
\caption{The V-band outburst light curves from St. Andrews, Keele and Essex. 
Light curves 85700 and 85703 from Keele have a time resolution of 3 seconds but 
they have been binned up by a factor of 4 for clarity. The drift from phase
zero of eclipse 85714 (observed from the JGT at St. Andrews) is believed to be
real.}
\label{fig:vband}
\end{figure}

The outburst light curves show extremely symmetric eclipses. This suggests 
that the disc is emitting more radiation than the white dwarf, the bright spot, 
and the secondary star combined. Evidence of this outburst being of an ``outside-in'' 
type comes from examining the individual R-band and V-band outburst light curves 
in Fig.~\ref{fig:rband} and Fig.~\ref{fig:vband} respectively. 

It can easily be seen that the rise to outburst
eclipses are wider than the eclipses in the late stages of the outburst (see Fig.~2, Fig~3 
and Fig.~6).Also, eclipse depths during the rise are much shallower than the eclipses during 
the decline, indicating that during mid eclipse there is more flux observed from either side 
of the secondary star during the rise phase than the decline phase of the outburst. 

Since the outburst eclipses are extremely symmetric, we fitted them with 
a Gaussian to measure the duration and depth of each eclipse. The residuals 
from the Gaussian fit were at the 5\% level. Table~\ref{tab:observation_log} lists the 
full width half minimum for each light curve along with the depth of each eclipse.
It is evident from Table~\ref{tab:observation_log} that the outburst eclipses
increase their depth and become narrower as the eruption progresses. It is a further 
indication that the outburst starts at the edge of the disc, whose luminous part is 
relatively large at the beginning of the outburst, and decreases in size as the outburst 
progresses. 

\scite{Patterson81} observed dwarf nova oscillations (DNO) during an eruption of HT
Cas in 1978. A power spectrum of our V and R out-of-eclipse data did not reveal any 
significant oscillations. Given that we have a relatively low temporal resolution 
and also that DNOs are usually found in shorter wavelength regions, this is not very 
surprising.

\section{Eclipse Timings and Ephemeris}

Table~\ref{tab:observation_log} shows the mid-eclipse timings obtained during 
the quiescent and outburst periods in 1995 and 1997. Previous data, that were
explicitly tabulated, from \scite{Patterson81}, \scite{Zhang86}, and 
Horne {\rm et~al.} (1991).
were converted from Heliocentric Julian Date to Barycentric Dynamical Julian Date and 
used in our calculations. Only the quiescent data were used in order to calculate the 
refined ephemeris presented here. The mid-eclipse positions were calculated by measuring 
the time of the white dwarf ingress and egress. 

For the outburst data presented here we used a Gaussian curve fit to measure the time of 
minimum flux. The data for November 17th from Keele Observatory have a timing problem and
they have been left out from all our calculations involving time and are not displayed on the 
O-C diagram.

Below is our calculated ephemeris in Barycentric Dynamical Julian Date with the 
uncertainties shown in brackets.

\begin{center}BDJD=2443727.93782(3)+0.073647211(5)$\times$E.\end{center}

\begin{figure}
\vspace{7.5cm}          

\includegraphics{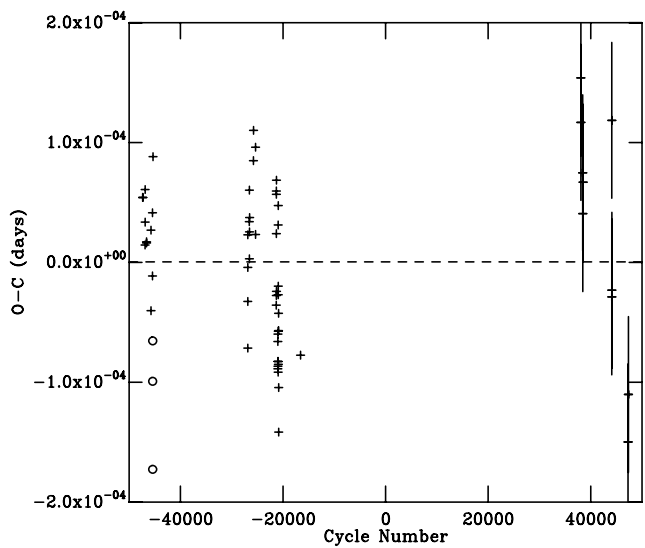}
\caption{$O-C$ diagram for HT Cas. The quiescent data are represented with
crosses. The outburst points that were observed by Patterson (1981) are
depicted as circles and were not used in our ephemeris calculations. The 
$T_{0}$ corresponds to BDJD=2447214.69133 and was placed in the middle of
our data range in order to give errors an equal leverage.}
\label{fig:o-c}
\end{figure}

We used our derived period to project back to the $T_{0}$ quoted by \scite{Patterson81}. 
The residuals of our ephemeris calculations (O-C diagram) can be seen in Fig.~\ref{fig:o-c}. 
The unused outburst eclipse points are represented as circles. \scite{Wood95} speculated that 
HT Cas might have a non zero period derivative. However, due to the large
scatter of the data points in Fig.~\ref{fig:o-c} we could only fit the data with
a linear ephemeris.
 
\begin{figure}
\vspace{7.5cm}          

\includegraphics{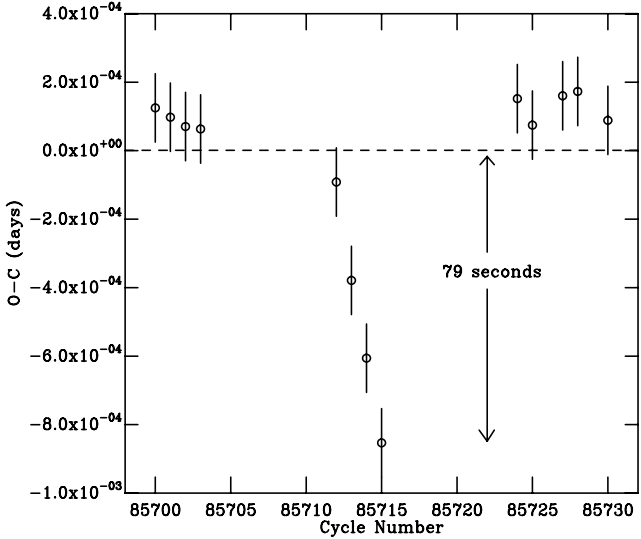}
\caption{The $O-C$ diagram for outburst eclipses of HT Cas during the 
November 1995 eruption. The large spread that was seen during the second night
of observations is difficult to explain.}
\label{fig:outburst_o-c}
\end{figure}

During the outburst an anomaly of the O-C points is observed, especially during the second night of 
observations when the system is at its early decline stages. (see Fig.~\ref{fig:outburst_o-c}) 
During the decline from outburst the points gradually drift to a maximum difference of 79 seconds 
from the calculated value. Observations of other objects on the same night show this is not 
a computer clock problem. The O-C points then return to their normal level the following night. 
We comment on this unusual behaviour in the discussion section of the paper.

\section{Modelling the Eclipse Data}
\label{mted}

A model that takes into account the flux emitted by a white dwarf encircled by
an optically thick accretion disc and a Roche-lobe filling secondary star, which 
is tidally locked in the frame of the binary, was modified and used for the investigation 
of the observed light curves during the HT Cas outburst. The model also accounts for 
the irradiation of the secondary's surface by a heating source at the centre 
of the disc. We also incorporate approximate temperature dependent limb darkening 
coefficients taken from \scite{Naimiy78}. Descriptions of previous versions of the 
code can be found in \scite{Shahbaz93} and \scite{Somers96a}. The model does not 
include the physical size and shape of the white dwarf or the white dwarf's shadow 
onto the accretion disc. Also, it takes no account of a bright spot or an 
accretion stream. 

It is evident from the symmetric outburst eclipses of HT Cas that flux 
from the erupting accretion disc completely dominates the flux contributions from 
the bright spot and the white dwarf, thus we believe that the assumption in the
modelling code of not considering flux from a bright spot is valid for this
case.

The overall temperature distribution of a dwarf nova disc that undergoes outbursts 
is clearly not that of a steady state model. However, at any time during the 
outburst, models for example by \scite{Pringle86}, show that the luminous parts 
of the disc undergoing an eruption closely resemble steady state models, but with 
truncated inner and outer radii. Thus we employed the standard steady state disc equation 
in order to calculate the flux from the disc, where the dependence of temperature is 
proportional to $(R_{wd}/R_{d})^{-3/4}$. We searched for an exponent for $R_{wd}/R_{d}$
that would reproduce the eclipse data better but we found that the preferred exponent
was always very close -- if not identical -- to $-3/4$, indicating that the
luminous part of the disc behaves in a quasi-steady state manner.

In our investigation of the eruption of HT Cas we were mostly concerned with the 
changes undergone by the accretion disc. Thus we fix those parameters  given in 
Table~\ref{tab:standard_param}. We apply a grid search for parameters that mainly 
concern the disc itself, for example the disc radius and the semi-opening angle. 

\begin{table}
\caption{List of standard parameters used to model the outburst eclipses of HT Cas, 
where $M_{1}$ mass of the primary, $u$ and ${\beta}$ are the limb and gravity darkening 
coefficients respectively and $T_{wd}$, $T_{pole}$ the white dwarf temperature
and the pole temperature of the secondary.}
\begin{tabular}{cll}
\hline

Parameter                &     Value   &    Reference    \\
\hline
Mass ratio $M_{2}/M_{1}$ & 0.18        & \scite{Catalan} \\

Period (days)            & 0.073647211 & This paper      \\

Inclination $(^\circ)$   & 80          & \scite{Catalan} \\

$M_{1} ({\Msun})$        & 0.91        & \scite{Catalan} \\

$u$                      & 0.71        & \scite{Naimiy78}\\

${\beta}$                & 0.08        & \scite{Sarna89} \\

Distance (pc)            & 140         & Wood {\rm et~al.} (1992)  \\

$T_{wd}$ (K)             & 14000       & Wood {\rm et~al.} (1992)  \\

$T_{pole}$ (K)           & 2500        & \scite{Baraffe96}\ \\

                         &             & \scite{Marsh90} \\                
\hline
\end{tabular}
\protect\label{tab:standard_param}
\end{table}

The parameters which concern the secondary star atmosphere have a very small, and 
often un-noticeable, effect on the overall light curve of the system. There is no 
indication of reprocessed radiation by the secondary star as the light curves do 
not show any orbital modulation that would be induced by irradiation from the white 
dwarf or a boundary layer. 

\begin{table*}
\caption{Results from modelling the outburst light curves of HT Cas.}
\begin{tabular}{lcccccccc}

q=0.18&Eclipse&\vline&Outer~$R_{d}$&Inner~$R_{d}$&Semi--opening&\.M&Reduced&fit\\
$(M_{2}/M_{1})$&Number&\vline&$(R_{d}/R_{L1})$&$R_{wd}$&angle$(^\circ)$&$({\Mdot})$&${\chi}^{2}$&\#\\
&&&&&${\pm}15\%$&${\pm}10\%$&\\
\hline
Rise to Outburst&(85700)&\vline&$0.41$&$4$&$10$&$9.3\times10^{-10}$&1.56&1\\
                &&\vline&$0.33$&0 (fixed)&$18$&$4.48{\times}10^{-10}$&3.56&2\\
                &&\vline&$0.63$&$3$&1 (fixed)&$1.59{\times}10^{-9}$&20.96&3\\
Peak of Outburst&(85703)&\vline&$0.37$&$1$&$15$&$6.9\times10^{-10}$&0.74&4\\
Early Decline&(85711)&\vline&$<0.41$&0&$11$&$4.5\times10^{-10}$&8.89&5\\
Late Decline&(85728)&\vline&$<0.32$&0&$9$&$1.3\times10^{-10}$&7.34&6\\

\hline
\end{tabular}
\protect\label{tab:Model_results}
\end{table*}

\begin{figure*}
\begin{minipage}{140mm}
\includegraphics{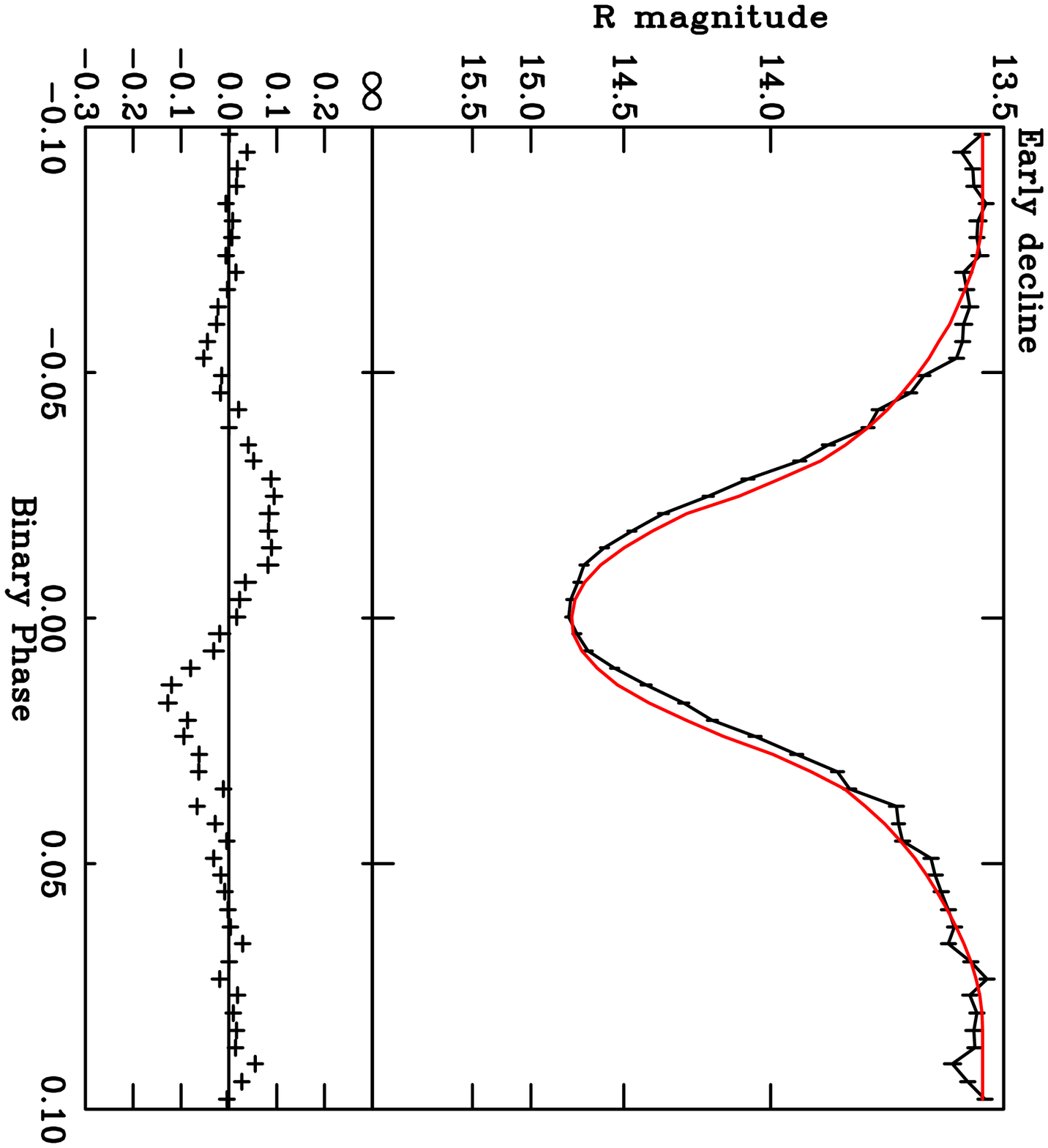}
\includegraphics{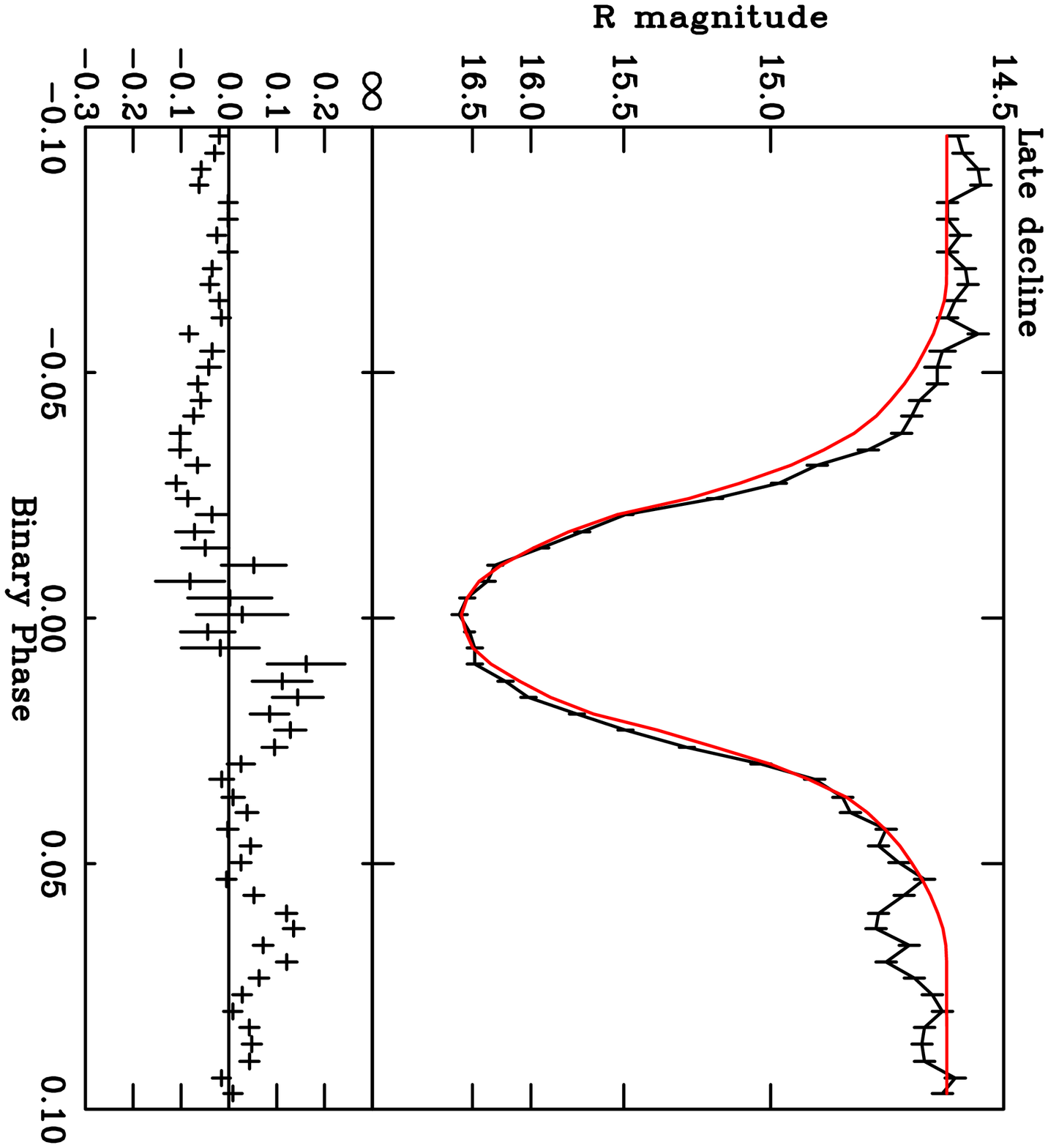}
\includegraphics{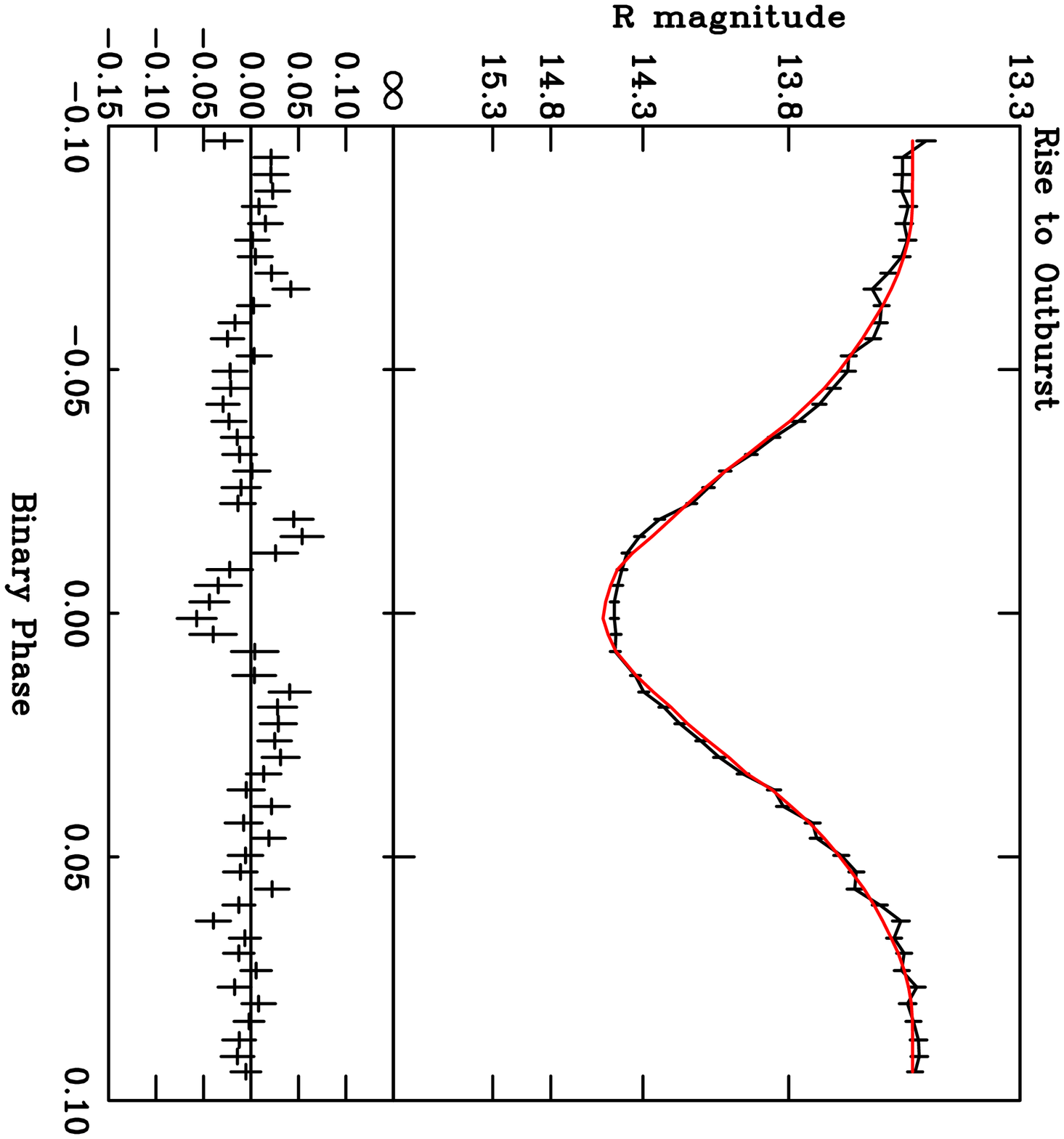}
\includegraphics{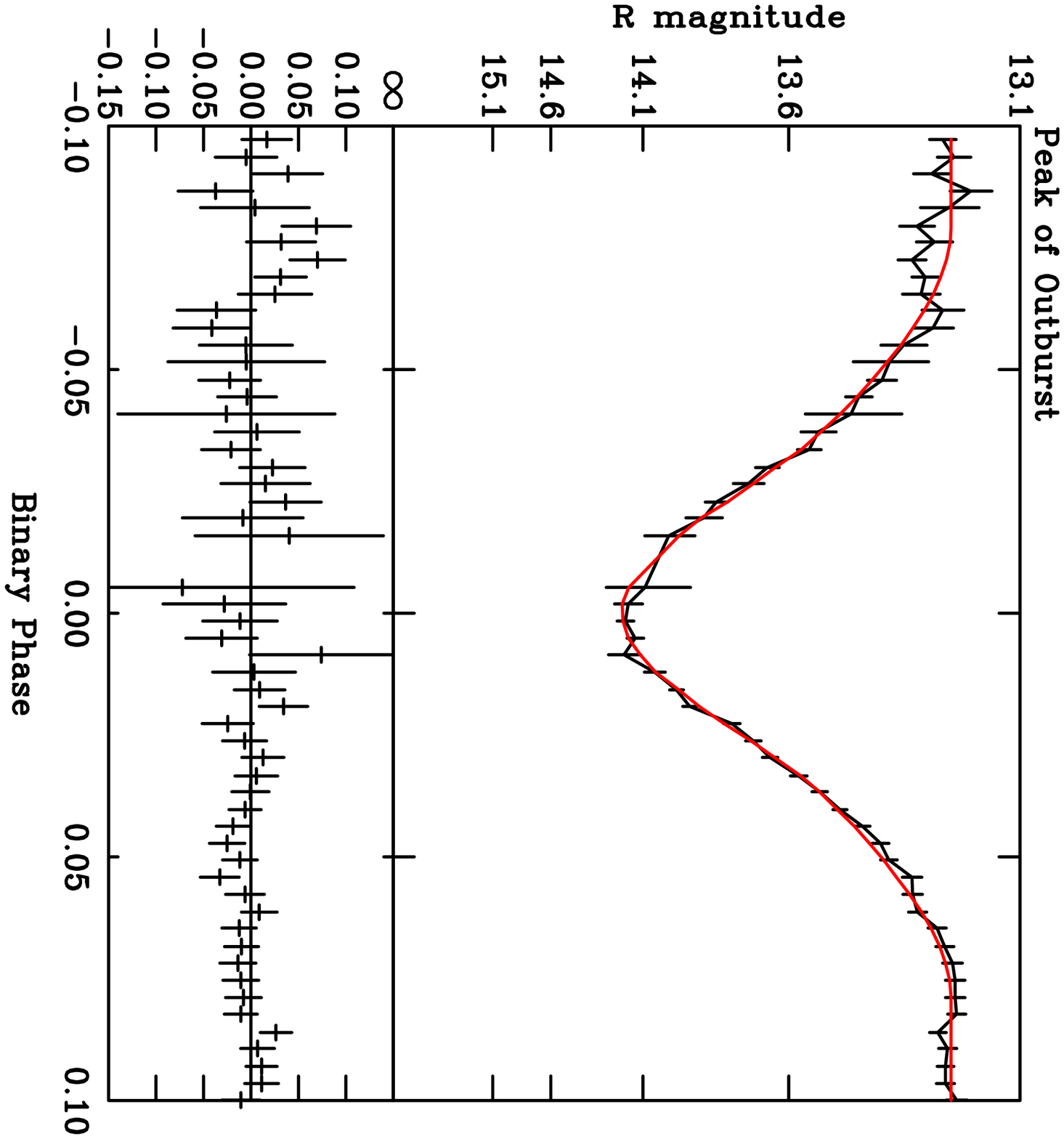}
\label{fig:resultgraphs}
\vspace{15cm}
\end{minipage}
\caption{Plot showing the model fits to the observed light curves from the HT Cas 
1995 outburst.}
\end{figure*}


\subsection{The rise--to--outburst eclipses}

All the fits presented here are models of the R-band data and are detailed in 
Table~\ref{tab:Model_results}. We cannot match the rise to outburst eclipses 
with a steady state model, even if we include considerable disc opening angles  
(Fit 2 in Table~\ref{tab:Model_results}). The eclipses that were generated by 
the model are always steeper than the data during ingress and egress. During 
eclipse minimum the steady state models could not reproduce the flatness shown 
by the data. 

To match the slope and depth of the eclipse light curves we used  
the code to calculate the flux from a truncated disc and made the code 
search for an inner truncation radius that would best match the 
data for each given outer radius. The semi-opening angle search in this case was 
coarser than the full disc models and varied from $1^{\circ}$ to $20^{\circ}$ in 
$2^{\circ}$ steps. 

The resulting models produced a much better fit to the data than the full disc models.  
The value returned by the model for the outer disc radius 
on the rise to outburst was $R_{d}=0.41R_{L1}$ and along with the value for the inner 
truncation radius of about 4~$R_{wd}$ and a flare angle of $10^{\circ}$ produced a fit 
with a ${\chi}^{2}_{\nu}$ of 1.56 (Fit 1 in Table~\ref{tab:Model_results}). We were still 
not able to reproduce exactly the flatness exhibited by the data at eclipse minimum, but 
ingress and egress are very well matched. Thus we conclude that in order to match the 
eclipses during the rising phase of the outburst, we need a truncated disc.

To prove that both the flare and truncation were required we ran a further set of models
(Fit 3), where the disc was truncated but not flared. The model profile did not match the 
observed data. The contact phases of the eclipse wings, which denote the contact phases of 
the disc edges with the secondary star extend beyond phases ${\pm}0.2$ and require a disc 
size of $R_{d}>0.8R_{L1}$. The ${\chi}^{2}_{\nu}$ for this model is very large with a value 
of 20.96 

The errors quoted for the semi-opening angle and the mass transfer rate in Table~\ref{tab:Model_results}
are empirical errors derived by fitting the data with models that lie above and
below the data points. 


\subsection{Eclipse colours on the rise--to--outburst}

\begin{figure}
\vspace{7.5cm}          

\includegraphics{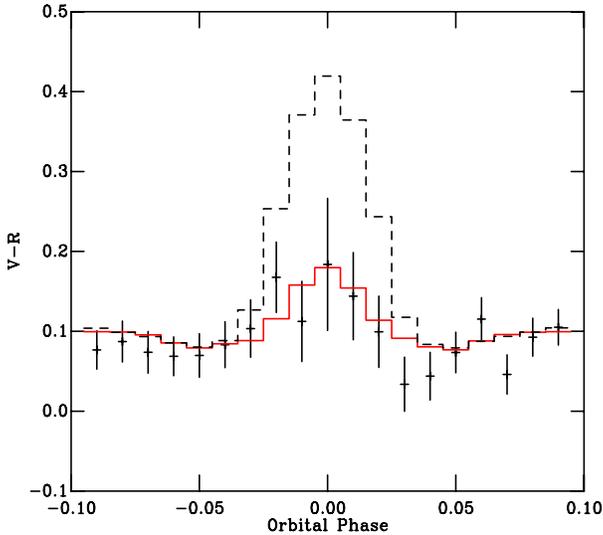}
\caption{Colour plot for eclipse 85700. Shown here are the binned $V-R$ light 
curve and the $V-R$ points that were calculated from the synthetic light 
curves. The solid line histogram depicts the $V-R$ of our best found model 
which includes a semi-opening angle of $10^{\circ}$ and a disc radius 
of $0.41R_{L1}$. The dotted histogram shows the $V-R$ behaviour of a large flat
disc (semi-opening angle of $1^{\circ}$ and $R_{d}>0.8R_{L1}$) without an 
inner truncation radius. It does not fit the observed data because the flux
distribution in this case is more centrally concentrated than our truncated
and flared disc, thus as the secondary star eclipses the inner parts of the
disc, a much more pronounced red feature is produced.}
\label{fig:v_r_1}
\end{figure}

The outburst eclipses 85700 and 85703 were observed in two colours. V-band
data were taken from Keele while the system was also observed in the R-band
from St. Andrews. Since the R-band data from St. Andrews were 
of lower temporal resolution, we binned the V-band data from Keele to match the 
exact phases of the St. Andrews data. The resulting $V-R$ eclipse light curve 
for cycle 85700 can be seen in Fig~\ref{fig:v_r_1}. For clarity the data
presented here were binned again into 0.01 phase bins starting from phase -0.1
to phase +0.1. 

The histograms in Fig~\ref{fig:v_r_1} are not model fits, but simply 
the $V-R$ calculated from the V and R synthetic light curves, created from the
best fit parameters for the $R$ band data. With a ${\chi}^{2}_{\nu}$ of 0.81 the 
solid line histogram in Fig~\ref{fig:v_r_1} is an adequate description to the data.
A constant of value 0.083 produced a ${\chi}^{2}_{\nu}$ of 1.24. Non-flared 
truncated disc models produced a $V-R$ fit with a ${\chi}^{2}_{\nu}$ of 1.51, 
but these models gave inadequate fits for the individual V and R eclipses 
(see Table~\ref{tab:Model_results}).

Thus we conclude that the system does not show any significant colour change 
during eclipse. To further emphasise the point that we cannot match the eclipses 
without a considerable flaring angle and without an inner truncation radius, 
we plot the $V-R$ values obtained from our best flat and non-truncated model disc. 
Clearly this model does not fit the data. 

\subsection{Peak of outburst}

For the eclipse close to the peak of the outburst the model prefers a smaller disc of 
radius $R_{d}=0.37R_{L1}$ with a larger flare angle of $15^{\circ}$ and also with the 
smaller inner truncation radius of about 1~$R_{wd}$. The ${\chi}^{2}_{\nu}$ for this 
model is 0.74 (Fit 4). 

\begin{figure}
\vspace{7.5cm}          

\includegraphics{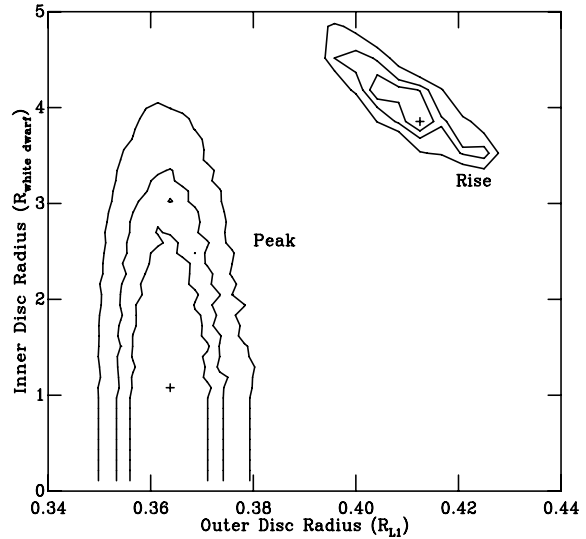}
\caption{Contour plot for the model outputs for the R-band eclipses 85700 and 85703. 
The y-axis represents the inner truncation radius required in terms of white dwarf 
radii from the white dwarf's surface. The crosses show the position of the minimum 
${\chi}^{2}$ found. The contours represent the 68\%, 90\% and 99\% confidence limits. 
In order to match the eclipses on the rise to outburst the model requires that the 
inner disc is truncated by ${\sim}4$ white dwarf radii for the outburst eclipse 85700 
(upper right region). For eclipse 85703 (larger lower left region) the contours do not 
constrain the truncation radius very well. This is because close to the peak of outburst 
the models prefer a high rim wall that hide the inner most parts of the disc.}
\label{fig:cont}
\end{figure}

With such a large flaring angle at the peak of outburst and given the 
$80^{\circ}$ inclination of this system, the inner parts on the far side of the accretion 
disc are hidden from view up to about 3$R_{wd}$ from the surface of the white dwarf. 
Thus we do not believe that the value of 1~$R_{wd}$ for the inner truncation
radius is physically meaningful. The contours for this eclipse (Fig.~\ref{fig:cont})
confirm this showing that the model is insensitive to values for the 
inner truncation radius up to about 3$R_{wd}$ from the white dwarf's surface.

The preference of the model to have a truncated disc on the rise to outburst 
suggests that either there is no material present in the inner most parts of 
the accretion disc (at least during the rising phase of the outburst) or that 
the hot transition front travelling from the outer parts of the disc towards the 
centre has not yet reached the inner parts of the accretion disc. In the second case, 
the inner disc is not void of material, but simply it does not emit significant amounts 
of flux because the hot transition front has not yet traversed the inner parts of the disc.

\subsection{Decline from outburst}

For the eclipses on the decline and late decline, the best models prefer a non--truncated 
disc, which show that the transition front, that originated at the outer parts of the disc 
and is traveling towards the centre has traversed the whole accretion disc, which has 
now started returning to its quiescent state (Fits 5 and 6).

At this stage the boundary layer around the white dwarf must have reached its
highest optical flux levels of the outburst, and the white dwarf must be
significantly hotter than during the rise. However, we do not observe any
effects of irradiation from the inner parts of the disc, the boundary layer, 
or the white dwarf onto the secondary star during any stage of the outburst.
This could be explained as the effect of a rim wall which prevents radiation from
the innermost parts of the disc reaching the secondary star surface. Also, effects 
of the irradiated secondary could be completely swamped due to the high flux levels
of the accretion disc.

During the decline from outburst there is a phase shift between the models and the 
actual data. The mid-eclipse in our models is always centered at phase zero, while 
the data seem to shift during the outburst, especially during the second night of 
observations. Regardless of the drift the eclipse profiles retain their high level 
of symmetry. Despite this peculiarity the model is able to reproduce the overall 
eclipse profile quite accurately. Undoubtedly, though, these models constitute only 
an upper limit on the disc radius and flare because the model uses a symmetric 
accretion disc and therefore any change of the eclipse ingress profile will similarly 
affect the egress profile. Indeed, when we manually shift the data to our 
calculated phase zero, the model eclipse profiles are in very good agreement with the 
observed data. The ${\chi}^2$ of the fits drops by a factor of 4 for the case of the 
early decline and also goes down by a factor of 2 for the modeled eclipse during the 
late stages of the decline. 

Although the values derived here for the semi-opening angle and the disc radius only 
constitute upper limits, they do show that the accretion disc is decreasing in size during 
the late stages of the decline.


\section{Discussion}

Investigation of the results presented in section~\ref{mted} show clearly that the
November 1995 outburst of HT Cas was of the ``outside - in'' type. 
We believe that the models show a relatively large luminous disc $R_{d}=0.41R_{L1}$ 
during the rise to outburst, which then shrinks to $(R_{d}<0.32R_{L1})$ in the later 
stages of the decline. It might be that the disc was smaller than $R_{d}=0.41R_{L1}$ 
at the early stages of the rise but unfortunately our coverage does not include 
observations at these outburst phases. 

Although our results seem to be in general agreement with the predictions of
the models of \scite{Mineshige85}, a discrepancy is noted. 
The quiescent disc radius for HT Cas was measured by Horne {\rm et~al.} (1991) 
and was found to be $R_{d}/{a}=0.23$ where a is the binary separation. This 
translates to $R_{d}/R_{L1}=0.33$ using a mass ratio of q=0.15. This value for 
$R_{d}$ is considerably less than our value of $R_{d}/R_{L1}=0.41$ during the 
rise of the outburst. Also, investigation of the SU UMa system Z Cha by 
\scite{O'Donoghue86} shows that there is a decrease in radius of the accretion 
disc in the time between outbursts. Assuming that HT Cas behaves in a similar 
manner this would mean that the pre-outburst disc was even smaller than that 
quoted by Horne {\rm et~al.} (1991). Thus the luminous radius of the disc on 
the rise to outburst is larger than the physical radius in quiescence.

This contrasts with the predictions of the models by \scite{Mineshige85}.
They show that a heating wave is initiated by a disc instability in the outer 
parts of the disc which travels inwards, with the luminous part never reaching 
the quiescent radius. This would suggest that during an outburst the disc would 
have a luminous size  roughly the same as that of the quiescent disc, with cool 
material on the outward side of the hot front expanding slightly in order to 
preserve angular momentum.

The disc never reaches the tidal radius, which for this system $(q=0.18)$ is at
$R_{tl}=0.76R_{L1}$. Thus, no superhump is observed during the eruption. Effects
in the $O-C$ diagram though are difficult to explain. During eruption, and in
particular during the second night of observations the $O-C$ values begin to
drift and reach up to a maximum delay of 79 seconds. The difficulty in explaining 
this feature as a simple dislocation of the flux centre comes from the fact that 
the eclipses retain their symmetry. Thus any dislocation of the flux centre must be 
accompanied by another effect, such as disc asymmetry, in order to counteract the 
change of flux centre and keep the overall eclipse shape symmetric.

The mass transfer rate through the disc, which is dependent on the distance, is 
found to be relatively high during the rise (\.M=$9.3\times10^{-10}{\Mdot}$).  
At the peak of the outburst the model results show a lower value of  
\.M=$6.9\times10^{-10}{\Mdot}$. This is surprising but it can be explained if 
we consider that when the system is at the peak of outburst the inner disc, unlike 
before, is now very hot and luminous. Therefore in order to match the flux level 
of the system we do not need high mass transfer rates. 

The \.M value quoted by \scite{Zhang86} for the January 1985 superoutburst is 
$3.8\times10^{-10}{\Mdot}$. This is about a factor of 2.5 times lower than our \.M 
estimate for the rise to outburst. Our slightly higher \.M values are easily explained 
by the fact that our disc models include an outer disc rim wall of semi-opening angle 
of $10^{\circ}$, which covers part of the disc and thus increasing the observed mass 
transfer rate necessary to match the flux level of the system. 

During the decline, we believe that a cooling front, as described in 
\scite{Mineshige85}, moves inward  reducing the disc temperature and at least 
for the duration of the eclipses the disc behaves in a steady state manner.
Also, \scite{Mineshige91} showed that for the case of Z Cha, for the disc 
temperature to decrease with time it is necessary for the disc to behave like, 
or at least be very close to, steady state. Conversely, models by \scite{Cannizzo94} 
on the decline of the optical fluxes during outbursts of dwarf novae require the 
accretion disc to be departing from steady state models. Clearly this is not the 
case that we observe, since our steady state models for the eclipses during the 
late decline are in good agreement with the observed data.

Comparing the results from the eruption of HT Cas with other outbursts of SU UMa
systems we see many similarities. In the case of OY Car, eclipse light curves 
observed by \scite{Vogt83} and later re--examined by \scite{Rutten92} show the same 
kind of behavior as the outburst eclipses of HT Cas. Both systems are slowly 
departing from their quiescent eclipse shape, moving towards a more symmetric shape 
with a distinct flat eclipse bottom during the early rise. These results are also
in extreme contrast with the results of \scite{Webb99} for the ``inside--out'' outburst
of the U~Gem system IP Peg.


\subsection{The inner disc ``truncation''}

A hot transition front of material, which is moving towards the centre of the disc, 
is suggested by the preference of the modelling code to place a hole 
(or non-luminous material) with a radius of about $4R_{wd}$ from the centre of
the accretion disc during the rising phase of the outburst. It is unclear as to 
whether the disc is truncated or not when at the peak of the outburst, as the 
flare angle of the disc grows to $15^{\circ}$ and so the innermost parts of the 
disc are hidden from view. During the decline from outburst the truncation radius 
is zero.

Several theories have been put forward in order to explain the delay in the rise
of the ultraviolet flux as opposed to the optical. For example, \scite{King97} 
proposed that the inner parts of the disc remain void of material during
quiescence due to irradiation from the white dwarf. Similarly \scite{Livio92} 
employed a weak magnetic field to truncate the disc, in the same manner as in 
intermediate polar systems. Also, \scite{Meyer90} suggested that a ``siphon'' 
coronal mass flow was responsible for emptying the inner parts of the accretion 
disc. Irradiation of the inner parts of the accretion disc by a hot corona that 
lies above the cool quiescent disc, leads to the evaporation of material in
those inner disc regions. Material from the disc joins the hot corona above and 
then is accreted onto the white dwarf leaving the inner parts of the accretion 
disc relatively empty. However, the fact that HT Cas shows X-ray eclipses
(Wood~{\etal}~1995; Mukai~{\etal}~1997) does not support the theory of a hot
X-ray producing corona, but instead suggests that the X-ray emitting regions in
HT Cas are very close to the white dwarf.

There has been previous observational evidence for the truncation of accretion
discs in non-magnetic dwarf nova systems. \scite{ladous96} found evidence of
truncated accretion discs for at least four different systems. They compared the
ratio of UV fluxes produced by white dwarfs and accretion discs as predicted
by the evaporation theory of Meyer and Meyer--Hofmeister \shortcite{Meyer89}
and \scite{Meyer90}.

In all three scenarios outlined above, the delay of the UV flux arises because of the 
time it takes for the inner parts of the disc to be filled by material in outburst 
that is moving inwards. Unfortunately we do not possess any UV data for this outburst 
in order to confirm that our model results are compatible with the above theories.
However, \scite{Wheatley96} and \scite{Naylor97} speculate that the delay in the UV 
flux observed in the VW Hyi system, which was observed simultaneously in the X-ray,
UV and optical parts of the spectrum during an eruption, could be explained by 
such a mechanism. Our data appear to show such a gap being filled.


\subsection{A vertically extended disc}

There is substantial evidence that the disc is highly flared throughout the
outburst, especially near the peak of the eruption.

The semi-opening angle of the disc is seen to expand from $10^{\circ}$ during the
rise to $15^{\circ}$ at the peak and then starts to decrease again during the
decline. This could possibly be explained by the ``avalanche'' effect that
\scite{Mineshige85} see in their model calculations. As the hot front
travels towards the inner parts of the disc, it causes the viscosity of the
material that has just gone through the heating front to rise sharply. Therefore
as the front moves inwards, material that is now in outburst, is piled up
behind the hot front perhaps thickening the disc. As the outburst declines, a cooling 
front that moves inwards makes the material at the outer parts of the disc return to 
their quiescent cold state. Again, as the cooling front travels inwards the disc is 
seen to decrease in size both radially and vertically.

Evidence that accretion discs thicken during outburst has been observed by others. 
In particular, in the case of another two SU UMa systems, OY Car and Z Cha. 
For the case of OY Car \scite{Naylor87} were confronted with a similar problem 
of not being able to fit the outburst eclipses with a steady state model and 
they speculate that this could be due to a rim wall shadowing the inner parts of 
the disc. Also, \scite{Naylor88} found evidence of a rim wall for OY Car by 
studying the UV and X--ray flux of the 1985 May superoutburst of OY Car. 
They suggest that the lack of an X--ray eclipse during the superoutburst, 
the existence of an orbital modulation of the UV flux and the observation that the 
UV spectra show emission lines which become narrower during eclipse, could be 
attributed to the fact that the processes taking place in the inner disc are partially 
hidden from view by a ``disc that has an extensive vertical structure''. In addition, 
\scite{Rutten92} found that by using the Maximum Entropy Mapping (MEM) method, in 
order to map the normal outburst eclipses of OY Car, they had to introduce an extra 
constant light source, which amounted to 15\% of the observed flux from the system. 
They speculated that this inconsistency could be attributed to the assumption of 
their model that the accretion disc height was negligible. 

One of the best pieces of observational evidence of flared accretion discs in SU UMa
systems during outburst, came from the investigation of the Z Cha system by \scite{Robinson95}. 
They observed in the ultraviolet using the {\it Hubble Space Telescope} and were able 
to follow the system during two eruption cycles. MEM models with negligible thickness 
produced an asymmetry creating a spurious ring in their disc maps due to extra 
flux that the MEM models could not account for. They found that in order to explain 
the extra flux observed they had to include a disc flare of $8^{\circ}$ into their MEM 
models. Blackbody curves were used to estimate the effective temperatures of the disc. 
This produces a temperature profile which was very flat and deviated significantly from 
profiles predicted by steady state models. They have recently \cite{Robinson99} improved 
their models using realistic model atmospheres in order to convert their surface brightnesses 
to effective temperatures, and also included the effect of limb darkening for accretion 
discs. They re-analysed their data for the Z Cha system and found that their temperature 
profile became steeper than their previous results, but it still did not match the steady 
state models. Their semi-opening angle for the accretion disc rim wall decreased but still 
remained considerable with a value of $6^{\circ}$.


\section{Conclusions}

Given the comprehensive coverage of the November 1995 outburst of HT Cas, spanning 
rise, maximum, and late decline, we were able to follow the changes induced 
to the light curve of HT Cas due to the erupting accretion disc. The four
significant results of this project are listed below.

(i)We have found that the outburst is initiated at the outer edges of the disc at 
a radius of $R_{d}=0.41R_{L1}$ and that the disc also expands vertically as the 
outburst evolves reaching a maximum of about $15^{\circ}$ when at the peak of
the outburst. These values of the disc flaring might not necessarily mean that
the accretion disc is flared to that extent, but they do emphasise that in order
to match the observed flux we really do need luminous material that extends
considerably above the orbital plane.

(ii)The models require disc semi-opening angles of the order of $10^{\circ}$ 
during the rise to outburst and $15^{\circ}$ close to the peak of outburst.
It is not possible to model the system during the rise or at the peak of
outburst simply by using conventional steady state models, or just flared disc models. 
In fact we have modified the model to search for an inner truncation radius for 
the disc and the results show that there might exist a heating front during 
the rise to outburst which moves towards the center of the disc. The size of the 
inner truncation radius seems to be about 4~$R_{wd}$ during the rise to outburst.

(iii)Using steady state models we do observe the mass transfer rate through the
disc to fluctuate by about a factor of 7 from the rise to outburst to late decline. 

(iv)On the decline, steady state models, do seem to agree with observational
data, although we detect a rather large deviation of the O-C points with the
eclipse shape retaining the high level of symmetry during these shifts.


\section{Acknowledgements}

We would like to thank Dr. Yvonne Unruh for acquiring the JGT data as well as
Prof. John Stull and amateur astronomer Mr. David Strange for supplementary data
and also Mr. Gary Poyner who contributed useful information concerning the
duration of the HT Cas 1995 outburst. Many thanks to the amateurs of Keele Astronomical 
Society for supporting and maintaining the Keele Observatory telescopes. Also, Dr. Janet 
Wood for useful comments and thoughts concerning this work. TN was supported by a PPARC 
advanced fellowship for part of this work. The data were reduced on the Keele
{\sc STARLINK} node using the {\sc ARK} software.
   


\begin{thebibliography}{{Ioannou, Naylor, Welsh, Catalan, Worraker, James}{1998}} 

\bibitem[\protect\citename{Al-Naimiy }1978]{Naimiy78}
Al-Naimiy~H.M.
1978, Ap.\&SS, 53, 181

\bibitem[\protect\citename{Baraffe \& Chabrier }1996]{Baraffe96}
Baraffe~I., Chabrier~G.,
1996, ApJ, 461, L51

\bibitem[\protect\citename{Bell, Hilditch \& Edwin }1993]{Bell93}
Bell~S.A., Hilditch~R.W., Edwin~R.P.,
1993, MNRAS, 260, 478

\bibitem[\protect\citename{Berriman, Kenyon \& Boyle }1987]{Berriman87}
Berriman~G., Kenyon~S., Boyle~C.,
1987, AJ, 94, 1291

\bibitem[\protect\citename{Catal\'{a}n }1995]{Catalan}
Catal\'{a}n~M.S.,
1995, Ph.D. Thesis, University of Sussex

\bibitem[\protect\citename{Cannizzo }1994]{Cannizzo94}
Cannizzo~J.K.,
1994, ApJ, 435, 389





\bibitem[\protect\citename{Harlaftis }1992]{Harlaftis92}
Harlaftis~E.T., Naylor~T., Hassall~B.J.M., Charles~P.A.,  Sonneborn~G.,
Bailey~J.,
1992, MNRAS, 259, 593

\bibitem[\protect\citename{Horne }1985]{Horne85}
Horne~K.,
1985, MNRAS, 213, 129

\bibitem[\protect\citename{Horne \& Cook }1985]{Horne85b}
Horne~K., Cook,M.C.,
1985, MNRAS, 214, 307

\bibitem[\protect\citename{Horne, Wood \& Stiening }1991]{Horne91}
Horne~K., Wood~J.H., Stiening~R.F.,
1991, ApJ, 378, 271



\bibitem[\protect\citename{King }1997]{King97}
King~A.R.,
1997, MNRAS, 288, L16

\bibitem[\protect\citename{La Dous {\rm et~al. }}1995]{ladous95}
La Dous~C., Meyer~F., Meyer--Hofmeister~E.,
1995, in Cataclysmic variables and related objects, IAU Coll. 158, Keele, eds.
A. Evans and J.H. Wood, Kluwer Astroph. and Space Science Library, p.81

\bibitem[\protect\citename{La Dous, Meyer \& Meyer--Hofmeister }1996]{ladous96}
La Dous~C., Meyer~F., Meyer--Hofmeister~E.,
1996, A\&A, 321, 213

\bibitem[\protect\citename{Livio \& Pringle }1992]{Livio92}
Livio~M., Pringle~J.E.,
1992, MNRAS, 259, 23

\bibitem[\protect\citename{Ludwig \& Meyer }1998]{Ludwig98}
Ludwig~K., Meyer~F.,
1998, A\&A, 329, 559

\bibitem[\protect\citename{Marsh }1990]{Marsh90}
Marsh~T.R.,
1990, ApJ, 357, 621

\bibitem[\protect\citename{Meyer }1990]{Meyer90}
Meyer~F., 
1990, Rev. Mod. Aston., 3, 1

\bibitem[\protect\citename{Meyer \& Meyer--Hofmeister }1984]{Meyer84}
Meyer~F., Meyer--Hofmeister~E.,
1984, A\&A, 132, 143

\bibitem[\protect\citename{Meyer \& Meyer--Hofmeister }1989]{Meyer89}
Meyer~F., Meyer--Hofmeister~E.,
1989, A\&A, 221, 36

\bibitem[\protect\citename{Mineshige \& Osaki }1985]{Mineshige85}
Mineshige~S., Osaki~Y.,
1985, Publs astr.Soc.Japan, 37, 1

\bibitem[\protect\citename{Mineshige }1991]{Mineshige91}
Mineshige~S., 
1991, MNRAS, 250, 253

\bibitem[\protect\citename{Misselt }1996]{Misselt96}
Misselt~K.A.,
1996, PASP, 108, 146

\bibitem[\protect\citename{Mukai {\rm et~al. }}1997]{Mukai97}
Mukai~K., Wood~J.H., Naylor~T., Schlegel~E.M., Swank~J.H.,
1997, ApJ, 475, 812

\bibitem[\protect\citename{Naylor {\rm et~al. }}1987]{Naylor87}
Naylor~T., Charles~P.A., Hassall~B.J.M., Bath~G.T., Berriman~G.,
Warner~B., Bailey~J., Reinsch~K.,
1987, MNRAS, 229, 183

\bibitem[\protect\citename{Naylor {\rm et~al. }}1988]{Naylor88}
Naylor~T., Bath~G.T., Charles~P.A., Hassall~B.J.M., Sonneborn~G.,
van der Woerd~H., van Paradijs~J.,
1988, MNRAS, 231, 237

\bibitem[\protect\citename{Naylor }1997]{Naylor97}
Naylor~T.,
1997, Ultraviolet Astrophysics Beyond The IUE Final Archive,
Sevilla, Spain, p.641

\bibitem[\protect\citename{Naylor }1998]{Naylor98}
Naylor~T.,
1998, MNRAS, 296, 339

\bibitem[\protect\citename{O'Donoghue }1986]{O'Donoghue86}
O'Donoghue~D.,
1986, MNRAS, 220, 23

\bibitem[\protect\citename{Osaki }1998]{Osaki96}
Osaki~Y.,
1996, PASP, 108, 39

\bibitem[\protect\citename{Patterson }1981]{Patterson81}
Patterson~J.,
1981, ApJS., 45, 517

\bibitem[\protect\citename{Pringle, Verbunt \& Wade }1986]{Pringle86}
Pringle~J.E., Verbunt~F., Wade~R.A.,
1986, MNRAS, 221, 169


\bibitem[\protect\citename{Ritter {\rm et~al. }}1998]{Ritter98}
Ritter~H., Kolb~U.,
1998, A\&AS, 129, 83

\bibitem[\protect\citename{Robinson {\rm et~al. }}1995]{Robinson95}
Robinson~E.L., Wood~J.H., Bless~R.C., Clemens~J.C., Dolan~J.F., Elliot~J.L.,
Nelson~M.J., Percival~J.W., Taylor~M.J., Van Citters~G.W., Zhang~E.,
1995, ApJ, 443, 295

\bibitem[\protect\citename{Robinson, Wood \& Wade }1999]{Robinson99}
Robinson~E.L., Wood~J.H., Wade~R.A,
1999, ApJ, in press

\bibitem[\protect\citename{Rutten {\rm et~al. }}1992]{Rutten92}
Rutten~R.G.M., Kuulkers~E., Vogt~N.,van~Paradijs~J.,
1992, A\&A, 265, 159

\bibitem[\protect\citename{Sarna }1989]{Sarna89}
Sarna~M.J.,
1989, A\&A, 224, 98

\bibitem[\protect\citename{Shahbaz, Naylor \& Charles }1993]{Shahbaz93}
Shahbaz~T., Naylor~T., Charles~P.A.,
1993, MNRAS, 265, 655

\bibitem[\protect\citename{Shahbaz }1994]{Shahbaz94}
Shahbaz~T.,
1994, Ph.D. Thesis, University of Keele

\bibitem[\protect\citename{Smak }1983]{Smak83}
Smak~J.,
1983, ApJ, 272, 234

\bibitem[\protect\citename{Somers, Mukai \& Naylor }1996a]{Somers96a}
Somers~M.W., Mukai~K., Naylor~T.,
1996a, MNRAS, 278, 845

\bibitem[\protect\citename{Somers {\rm et~al. }}1996b]{Somers96b}
Somers~M.W., Lockley~J.J., Naylor~T., Wood~J.H.,
1996b, MNRAS, 280, 1277



\bibitem[\protect\citename{Vogt }1983]{Vogt83}
Vogt~N.,
1983, A\&A, 128, 29

\bibitem[\protect\citename{Wade }1984]{Wade84}
Wade~R.A.,
1984, MNRAS, 208, 381

\bibitem[\protect\citename{Warner }1995]{Warner95}
Warner~B.,
1995, Cataclysmic variable stars, Cambridge University Press.

\bibitem[\protect\citename{Webb {\rm et~al. }}1999]{Webb99}
Webb~N.A., Naylor~T., Ioannou~Z., Worraker~W.J., Stull~J., Allan~A.,
Fried~R., James~N.D, Strange~D.,
1999, in preparation

\bibitem[\protect\citename{Wheatley {\rm et~al. }}1996]{Wheatley96}
Wheatley~P.J., Verbunt~F., Belloni~T., Watson~M.G., Naylor~T., Ishida~M.,
Duck~S.R., Pfeffermann~E.,
1996, A\&A, 307, 137

\bibitem[\protect\citename{Wlodarczyk }1986]{Wzyk86}
Wlodarczyk~K.,
1986, Acta Astr., 36, 395 

\bibitem[\protect\citename{Wood, Horne \& Vennes }1992]{Wood92}
Wood~J.H., Horne~K., Vennes~S.,
1992, ApJ, 385, 294

\bibitem[\protect\citename{Wood {\rm et~al. }}1995]{Wood95}
Wood~J.H., Naylor~T., Hassall~B.J.M., Ramseyer~T.F.,
1995, MNRAS, 273, 772


\bibitem[\protect\citename{Young, Schneider \& Shectman }1981]{Young81}
Young~P., Schneider~D.P., Shectman~S.A.,
1981, ApJ, 245, 1035

\bibitem[\protect\citename{Zhang, Robinson \& Nather }1986]{Zhang86}
Zhang~E.H., Robinson~E.L., Nather~R.E.,
1986, ApJ, 305, 740

\end{thebibliography}
\end{document}